\documentclass[twocolumn,superscriptaddress]{revtex4}
\usepackage{mathrsfs}
\usepackage{amssymb,amsbsy,amsmath,latexsym,dsfont,array,layout,graphicx,mathrsfs,color}
\usepackage{multirow}
\usepackage{footnote}
\usepackage{setspace}
\usepackage{subfigure}

\newcommand{\ket}[1]{\left|{#1}\right\rangle}
\newcommand{\bra}[1]{\left\langle{#1}\right|}

\begin{document}

\title{Arbitrary coherent distributions in a programmable quantum walk}
\author{Rong Zhang}\altaffiliation{These authors contributed equally to this work.}
\affiliation{National Laboratory of Solid State Microstructure, School of Physics, School of Electronic Science and Engineering, and Collaborative Innovation Center of Advanced Microstructures, Nanjing University, Nanjing, 210093, China}
\affiliation{College of Electronic and Optical Engineering, Nanjing University of Posts
and Telecommunication, Nanjing 210023, China}

\author{Ran Yang}\altaffiliation{These authors contributed equally to this work.}
\affiliation{National Laboratory of Solid State Microstructure, School of Physics, School of Electronic Science and Engineering, and Collaborative Innovation Center of Advanced Microstructures, Nanjing University, Nanjing, 210093, China}

\author{Jian Guo}
\affiliation{National Laboratory of Solid State Microstructure, School of Physics, School of Electronic Science and Engineering, and Collaborative Innovation Center of Advanced Microstructures, Nanjing University, Nanjing, 210093, China}

\author{Chang-Wei Sun}
\affiliation{National Laboratory of Solid State Microstructure, School of Physics, School of Electronic Science and Engineering, and Collaborative Innovation Center of Advanced Microstructures, Nanjing University, Nanjing, 210093, China}

\author{Yi-Chen Liu}
\affiliation{National Laboratory of Solid State Microstructure, School of Physics, School of Electronic Science and Engineering, and Collaborative Innovation Center of Advanced Microstructures, Nanjing University, Nanjing, 210093, China}

\author{Heng Zhou}
\affiliation{School of Information and Communication Engineering, University of Electronic Science and Technology of China, Chengdu 611731, China}

\author{Ping Xu}\email{pingxu520@nju.edu.cn}
\affiliation{Institute for Quantum Information and State Key Laboratory of High Performance Computing, College of Computing, National University of Defense Technology, Changsha, 410073, China}
\affiliation{National Laboratory of Solid State Microstructure, School of Physics, School of Electronic Science and Engineering, and Collaborative Innovation Center of Advanced Microstructures, Nanjing University, Nanjing, 210093, China}

\author{Zhenda Xie}\email{xiezhenda@nju.edu.cn}
\affiliation{National Laboratory of Solid State Microstructure, School of Physics, School of Electronic Science and Engineering, and Collaborative Innovation Center of Advanced Microstructures, Nanjing University, Nanjing, 210093, China}
\author{Yan-Xiao Gong}\email{gongyanxiao@nju.edu.cn}

\affiliation{National Laboratory of Solid State Microstructure, School of Physics, School of Electronic Science and Engineering, and Collaborative Innovation Center of Advanced Microstructures, Nanjing University, Nanjing, 210093, China}

\author{Shi-Ning Zhu}
\affiliation{National Laboratory of Solid State Microstructure, School of Physics, School of Electronic Science and Engineering, and Collaborative Innovation Center of Advanced Microstructures, Nanjing University, Nanjing, 210093, China}


\begin{abstract}
  The coherent superposition of position states in a quantum walk (QW) can be precisely engineered
  towards the desired distributions to meet the need of quantum information applications.
  The coherent distribution can make full use of quantum parallel in computation and simulation.
  Particularly, the uniform superposition provides the robust non-locality,
  which has wide applications such as the generation of genuine multi-bit random numbers without post-processing.
  We experimentally demonstrate that the rich dynamics featured with arbitrary coherent distributions
  can be obtained by introducing different sets of the time- and position-dependent operations.
  Such a QW is realized by a resource-constant and flexible optical circuit, in which the variable operation is executed
  based on a Sagnac interferometer in an intrinsically stable and precisely controlled way.
  Our results contribute to the practical realization of quantum-walk-based quantum computation, quantum simulations and quantum information protocols.
\end{abstract}

\maketitle
\section{INTRODUCTION}
Classical random walk (CRW) is a basic model of stochastic processes with applications
in computation science~\cite{MR95}, and quantum walk (QW) is a generalization of CRW to
the quantum regime~\cite{AZ93}. Due to the coherent superposition and quantum interference,
QW outperforms CRW in computation by intriguing quantum algorithms~\cite{AC09}.
In the discrete-time QW, the evolution of a quantum particle depends on the state of a coin,
and hence controlling the coin degree of freedom can indirectly engineer the QW dynamics
featured with various walker states~\cite{VA12}.
So QW with diverse types of coin operations has been explored theoretically and experimentally
to carry out quantum information protocols~\cite{TG19,VR13,WG18,KW13,BX15,SS11,SS12,KW12,XX17,BS18,CC18,MM16,MM17}.
One type is the time-dependent coin operation which has been employed to
produce high-dimensional quantum states~\cite{TG19}
and the maximal entanglement~\cite{VR13,WG18}. The other type is the position-dependent coin operation which
has been utilized to realize the general quantum measurement~\cite{KW13,BX15} and
simulate complex phenomena~\cite{SS11,SS12,KW12,XX17,BS18,CC18}. Furthermore, a controllable arrangement of
both time- and position-dependent coin operations is used to enhance the degree of control
over the quantum walker's dynamics, thereby more versatile dynamics and coherent superposition
with arbitrary distributions have been investigated theoretically~\cite{MM16,MM17}.
However, for experimental realization, such manipulation requires stably and precisely
controlled way to engineer the coin operations at each step and each position,
which has not been realized in any physical system yet.

The coherent superposition of many position states is of great importance in quantum information science and technology. 
For instance, the coherent uniform superposition can be used to
generate the multi-bit random number by exploiting the intrinsic unpredictability, and the rate of random bits
can be readily enhanced by the increased evolution time of the well-engineered QW~\cite{GS14,MQ16,MC17,SC19}.
Moreover, the uniform distribution has been extensively employed in efficient algorithms~\cite{MR95}. And another important
example is the coherent Gaussian distribution. As CRWs exhibit Gaussian distribution,
QWs with Gaussian distribution can simulate the dynamics of classical systems~\cite{MM16}. 
In comparison to QWs with incoherent states by introducing de-coherence, random operations,
or measurements~\cite{BA03,XS13,KT03}, QWs with coherent distributions~\cite{MM16,MM17} evolve
in a unitary way and can make full use of quantum parallel in computation and simulation.

Here we present the realization of arbitrary coherent distributions in the programmable QW
by employing a time-bin encoded optical loop configuration.
We confirm the dynamic richness of the QW by observing the ballistic distribution, the uniform and Gaussian distributions with well-preserved coherence.
The coherence during the whole evolution can be confirmed if the final states are pure.
Inspired by~\cite{SS11,SS12,BS18,NS18,RP11,NS16,BW16,HS16,LB19,ES15}, but,
instead the time- and position-dependent coin operations is implemented
by utilizing a stable polarization Sagnac interferometer.
Due to the well-established equivalence between the evolution of coherent light in a linear optical network
and that of a single photon [see p. 106 in ~\cite{HP04}],
we use the coherent input light to experimentally simulate the versatile dynamics of single photons.
Combining the diverse initial state preparation, our programmable QW setup, which
allows for the full control of the coin operation,
can be used to create coherent superposition of many position states as desired definitely and to investigate versatile interaction dynamics showing superiority
for applications in quantum computation, quantum simulations, and other quantum information technologies.~\cite{YA21,EA21}

\section{Arbitrary coherent distributions}
The one-dimensional discrete-time QW involves the coin and the walker described in the tensor product of two subspaces $\mathscr{H}_\text{c}\otimes\mathscr{H}_\text{w}$.
The basis state can be expressed as $\ket{c,x}$, with $c = 0, 1$ representing two coin states, and $x\in \mathbb{Z}$ the discrete positions of the walker, respectively. Here we employ the real-valued time- and position-dependent coin operation~\cite{MM16,MM17} expressed as
\begin{equation}
\label{eq:C}\hat{C}_{x,t}=
\begin{pmatrix}
        \text{cos}\theta_{x,t} & \text{sin}\theta_{x,t}\\
        \text{sin}\theta_{x,t} & -\text{cos}\theta_{x,t}
\end{pmatrix},
\end{equation}
with the parameter $\theta_{x,t} \in [0, \pi]$ depending on the evolution time $t$
and the position $x$. The QW evolution is described by the unitary operator
\begin{equation}
\label{eq:U}\hat{U}_{t}=\sum_x \hat{S}_x \left[\hat{C}_{x,t}\otimes \hat{I}\right],
\end{equation}
where $\hat{I}$ is the identity operator on the positions of walker,
and the shift operation $\hat{S}_x$ makes the walker move to the neighboring left (right) position
according to the coin state $\ket{0}$ ($\ket{1}$), which is
expressed as $\hat{S}_x=\ket{0}\bra{0}\otimes\ket{x-1}\bra{x}+\ket{1}\bra{1}\otimes\ket{x+1}\bra{x}$.

The initial coin-walker state is $\psi(t=0)=[a(0,0)\ket{0}+b(0,0)\ket{1}]\ket{0}$, with
$\left|a(0,0)\right|^2+\left|b(0,0)\right|^2=1$. For simplicity, $a(0,0)$ and $b(0,0)$
are assumed to be real valued, however, the method can be generalized to the complex-valued
case as well. Then after $t$-step evolution given by
$\psi(t)=\hat{U}_{t-1}\cdots\hat{U}_1\hat{U}_0\psi(0)$, the final state is
\begin{equation}
\label{eq:Ft} \psi(t)=\sum_x \left[a(x,t)\ket{0}+b(x,t)\ket{1}\right]\ket{x},
\end{equation}
with $x = -t,-t+2, \cdots, t$. All amplitudes of $\psi(t)$ can be derived from the following recursive equation
\begin{align}
\label{eq:x+1}a(x+1,t+1)=&cos\theta_{x,t}a(x,t)+ sin\theta_{x,t}b(x,t),\\
\label{eq:x-1}b(x-1,t+1)=&sin\theta_{x,t}a(x,t)- cos\theta_{x,t}b(x,t).
\end{align}
Eqs.~(\ref{eq:x+1})-(\ref{eq:x-1}) fully determine the dynamics of the single particle
in the QW once the initial state is set.

It has been proved that the QW with time- and position-dependent coin operations can
be designed to obtain the desired evolution featured with a fixed probabilistic function
$P(x,t)$, which gives the likelihood of the particle at position $x$ after $t$-step evolution~\cite{MM16,MM17}.
Due to the real valued coefficients of both initial states and evolution operators,
all amplitudes of the final state are real-valued. Now the position probability distribution
is $P(x,t)=a^2(x,t)+b^2(x,t)$. The arbitrarily coherent superposition state with a fixed $P(x,t)$ can be achieved by employing the sound set of
time- and position-dependent coin operations~\cite{MM17}
\begin{align}
\label{eq:cos}cos\theta_{x,t}=&\frac{a(x,t)a(x+1,t+1)}{a^2(x,t)+b^2(x,t)}-\frac{b(x,t)b(x-1,t+1)}{a^2(x,t)+b^2(x,t)},\\
\label{eq:sin}sin\theta_{x,t}=&\frac{b(x,t)a(x+1,t+1)}{a^2(x,t)+b^2(x,t)}+\frac{a(x,t)b(x-1,t+1)}{a^2(x,t)+b^2(x,t)}.
\end{align}
It is noted that the initial state $\psi(0)$ can be arbitrary provided that the proper $cos\theta_{0,0}$ and $sin\theta_{0,0}$ are chosen~\cite{MM17}.

Two illustrated examples, the coherent Gaussian~\cite{MM16} and uniform~\cite{MM17} distributions, are given here.
The Gaussian distribution is described by $P^G(x,t)=\frac{1}{2^t}\frac{t!}{\left(\frac{t-x}{2}\right)!\left(\frac{t+x}{2}\right)!}$
with $x=-t, -t+2, \cdots, t$.
For our realization, the initial state is $\psi(0)=\ket{0}_{\text{c}}\ket{0}_{\text{w}}$,
so $cos\theta_{0,0}=sin\theta_{0,0}=\frac{1}{\sqrt{2}}$ are chosen.
The matrix elements for $t \ge 1$ are
\begin{align}
\label{eq:cosg}cos\theta^G_{x,t}=&\frac{1}{2}\left(\sqrt{1+\frac{x}{t}}-\sqrt{1-\frac{x}{t}}\right),\\
\label{eq:sing}sin\theta^G_{x,t}=&\frac{1}{2}\left(\sqrt{1+\frac{x}{t}}+\sqrt{1-\frac{x}{t}}\right).
\end{align}
After $t$-step evolution of Eqs.~(\ref{eq:C}), (\ref{eq:U}), (\ref{eq:cosg}) and (\ref{eq:sing}), the updated state is $\psi^G(t)=\sum_x\left[a_G(x,t)\ket{0}+b_G(x,t)\ket{1}\right]\ket{x}$, with $a^2_G(x,t)+b^2_G(x,t)=P^G(x,t)$.
To remove correlations between coin and walker and further get the pure walker state with Gaussian distribution $\psi^{G}_x(t)=\sum_x\sqrt{P^G(x,t)}\ket{x}$,
the coin operations
$\frac{1}{N^G_{x,t}}\begin{pmatrix}
        a_G(x,t) & b_G(x,t)\\
        b_G(x,t) & -a_G(x,t)
\end{pmatrix}$
with $N^G_{x,t}=\sqrt{a^2_G(x,t)+b^2_G(x,t)}$ are executed on the state $\psi^G(t)$, and the state $\psi^{G'}(t)=\ket{0}\psi^{G}_x(t)$ is achieved.
Thereby we can easily get the coherent superposition state
$\psi^{G}_x(t)$ with the Gaussian distribution.
Therefore, the quantum walker suffers unitary dynamics in a deterministic and reversible way and exhibits the binomial distribution the same with CRW.

By choosing another set of coin operations, the walker can be prepared in the equally coherent
superposition of position states leading to the uniform distribution $P^u(x,t)=\frac{1}{t+1}$
with $x=-t, -t+2, \cdots, t$~\cite{MM17}.
The initial state is also $\ket{0}_{\text{c}}\ket{0}_{\text{w}}$ and
$cos\theta_{0,0}=sin\theta_{0,0}=\frac{1}{\sqrt{2}}$ are chosen. For $t \ge 1$ the matrix elements are
\begin{align}
\label{eq:cosu}cos\theta^u_{x,t}=&\frac{1}{2}\sqrt{\frac{(t+x)(t+x+2)}{t(t+2)}}-\frac{1}{2}\sqrt{\frac{(t-x)(t-x+2)}{t(t+2)}},\\
\label{eq:sinu}sin\theta^u_{x,t}=&\frac{1}{2}\sqrt{\frac{(t+x)(t+x+2)}{t(t+2)}}+\frac{1}{2}\sqrt{\frac{(t-x)(t-x+2)}{t(t+2)}}.
\end{align}
After $t$-step evolution described by Eqs.~(\ref{eq:C}), (\ref{eq:U}), (\ref{eq:cosu}) and (\ref{eq:sinu}),
the evolved state is $\psi^u(t)=\sum_x\left[a_u(x,t)\ket{0}+b_u(x,t)\ket{1}\right]\ket{x}$
with $a^2_u(x,t)+b^2_u(x,t)=\frac{1}{t+1}$. Then by employing
the coin operations $\frac{1}{N^u_{x,t}}\begin{pmatrix}
        a_u(x,t) & b_u(x,t)\\
        b_u(x,t) & -a_u(x,t)
\end{pmatrix}$ with $N^u_{x,t}=\sqrt{a^2_u(x,t)+b^2_u(x,t)}$, the state $\psi^{u}(t)$ evolves into $\psi^{u'}(t)=\frac{1}{\sqrt{t+1}}\ket{0}\sum_x\ket{x}$. Thus we definitely obtain the equally coherent superposition
of all possible position states
\begin{equation}
\label{eq:uni1}\psi^{u}_x(t)=\frac{1}{\sqrt{t+1}}\sum_x\ket{x},
\end{equation}
with $x=-t, -t+2, \cdots, t$. Furthermore, with $\ket{0}$ ($\ket{1}$) representing zero (one) particle at the position $x$,
the walker's state $\psi^{u}_x(t)$ can also be expressed as the quantum state of $(t+1)$ qubits~\cite{GS14}
\begin{equation}
\label{eq:uni2}\psi^{u}_{\text{W}}(t)=\frac{1}{\sqrt{t+1}}\left(\ket{10\cdots0}+\ket{01\cdots0}+\cdots+\ket{0\cdots01}\right).
\end{equation}
$\psi^{u}_{\text{W}}(t)$ is the maximal high order W-type entanglement which behave the robust non-locality and have wide applications~\cite{GS14}.

The genuine multi-bit random number can be generated by exploiting the intrinsic uncertainty
in the coherent uniform distribution expressed by Eq.~(\ref{eq:uni1}) or Eq.~(\ref{eq:uni2}).
For example, after $7$-step evolution, the walker is in the equally coherent superposition of
eight positions $x \in [-7, -5, \cdots, 7]$, and
then a $3$-bit random number can be produced after a single quantum measurement.
The bit generation rate can be readily improved by increasing the evolution steps of appropriately engineered QW~\cite{SC19}.
The multi-bit method provides the advantage of a larger attainable bit rate for a given particle flux~\cite{GS14},
in compared with the single bit schemes with limited bit rat due to detector saturation.
Furthermore, the method presented here only requires the conversion from particle counting to the digital format
without the need of post-processing.

\section{Experimental realization}
We realize the programmable QW 
by introducing the precisely controlled time- and position-dependent coin operations as shown in Fig.~1.
We use the horizontal polarization $\ket{H}$ and the vertical one $\ket{V}$
of light to represent the two coin states $\ket{0}$ and $\ket{1}$, respectively.
And we transform the positions of the walker $\ket{x}$ with $x=-t, -t+2, \cdots, t$
into the arrival times at the detector by routing the initial light pulse to paths of different length.
The input light is generated by a pulsed diode laser with a central wavelength $1550$~nm, a pulse
width of $70$~ps, and a repetition of $1$~MHz, which is triggered by the
electrical pulse output from arbitrary wave generator (AWG) (Tektronix, AWG7012).
The light pulses are attenuated to the single-photon level by using the combination of
the polarization beam splitter (PBS0), half-wave plate (HWP0) and PBS1. 
The initial local state $\psi(0)$ is prepared
by  HWP1 or QWP1 after PBS1. Then the light is coupled into the QW
network via the $1/99$ port of a $1:99$ single-mode fiber beam splitter (BS1).

\begin{figure}[t]
        \includegraphics[width=0.5\textwidth]{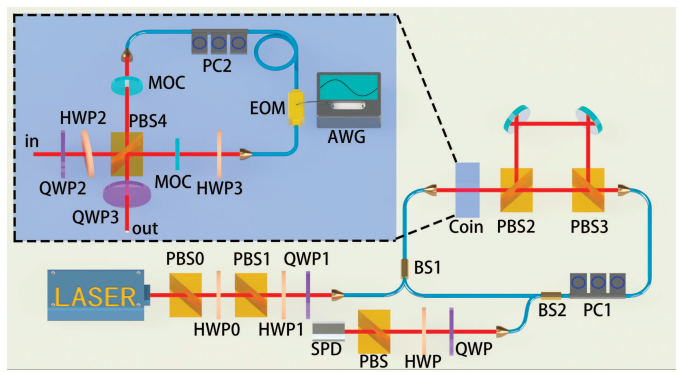}
        \caption{Experimental setup of the time- and position-inhomogeneous quantum walk. PBS: polarization beam splitter; PBS: polarization beam splitter; HWP: half-wave plate; QWP: quarter-wave plate; BS: single-mode fiber beam splitter; PC: polarization controller; MOC: magneto-optic crystal; EOM: electro-optic phase modulator; AWG: arbitrary wave generator; SPD: single-photon detector.
    }\label{Fig:var-exp}
\end{figure}

The variable coin operation of Eq.~(1) can be decomposed into
\begin{equation}
\label{eq:Cexp}\hat{C}_{x,t}=\frac{1}{\sqrt{2}}
\begin{pmatrix}
        1 & -i\\
        -i & 1
\end{pmatrix}
\begin{pmatrix}
        e^{i\phi_H(x,t)} & 0\\
        0 & e^{i\phi_V(x,t)}
\end{pmatrix}
\begin{pmatrix}
        1 & -i\\
        -i & 1
\end{pmatrix},
\end{equation}
with the phase modulation of the horizontal polarization
$\phi_H(x,t)=\theta_{x,t}$ and the vertical one $\phi_V(x,t)=\pi-\phi_H(x,t)$.
So the coin operation is realized by the sandwich scheme
QWP2 + Sagnac loop + QWP3
as shown in the part surrounded by the dashed line in Fig.~1.
QWP2 and QWP3 are both aligned at $\pi/8$, and
the Sagnac loop is used to tune the phase $\phi_{H/V}(x,t)$ precisely
because the phase modulations generated directly by the electro-optic
phase modulator (EOM) (Eospace) depend on $H$ and $V$ polarizations with the restriction $\phi_V/\phi_H=3.5$.
In the Sagnac interferometer, firstly, the light pulse is separated spatially by PBS4 depending on polarization,
and then H and V polarized components are guided into counter-clockwise and clockwise paths, respectively.
Secondly, the arrival time at the EOM of the clockwise light pulse is delayed $39.1$~ns
compared with that of the counter-clockwise one by fusing a single-mode fiber
with the upper polarization-maintaining (PM) fiber of the EOM, so that the phases of H and V polarization can be precisely modulated
by applying the voltages on the EOM independently.
Thirdly, A magneto-optic crystal (MOC) is placed in each path to make the polarization rotate $\pi/4$
counter-clockwise.
Both ports of the EOM are connected with PM
fibers aligned in the slow axis. The optical axis of HWP3 is tuned to make the
diagonal polarization aligned with the slow axis of the PM fiber. Likewise, the polarization controller
PC2 is adjusted to align the anti-diagonal polarization with the slow axis of the PM fiber.
The tilting HWP2 is used to compensate the intrinsic different phase between the clockwise and
counter-clockwise light of the Sagnac loop.
Finally, the EOM is triggered by programmable electric pulses output from the AWG (Tektronix, AWG7012),
and then the light pulses of both clockwise and counter-clockwise path return to the original polarization at the output of
the Sagnac interferometer with an individual modulated phase $\phi_{H/V}(x,t)$.

\begin{figure}[t]
    \includegraphics[width=0.48\textwidth]{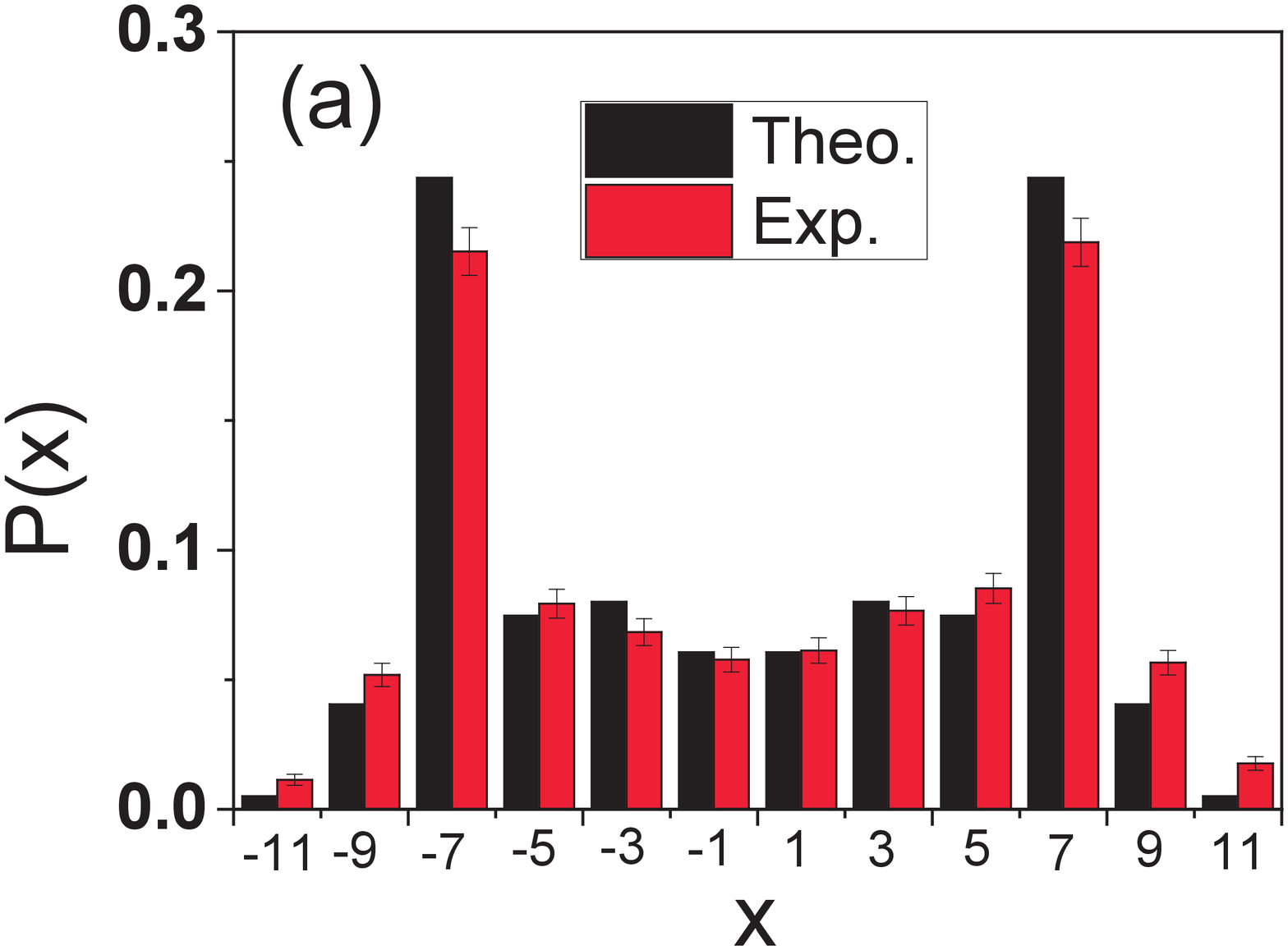}
    \includegraphics[width=0.48\textwidth]{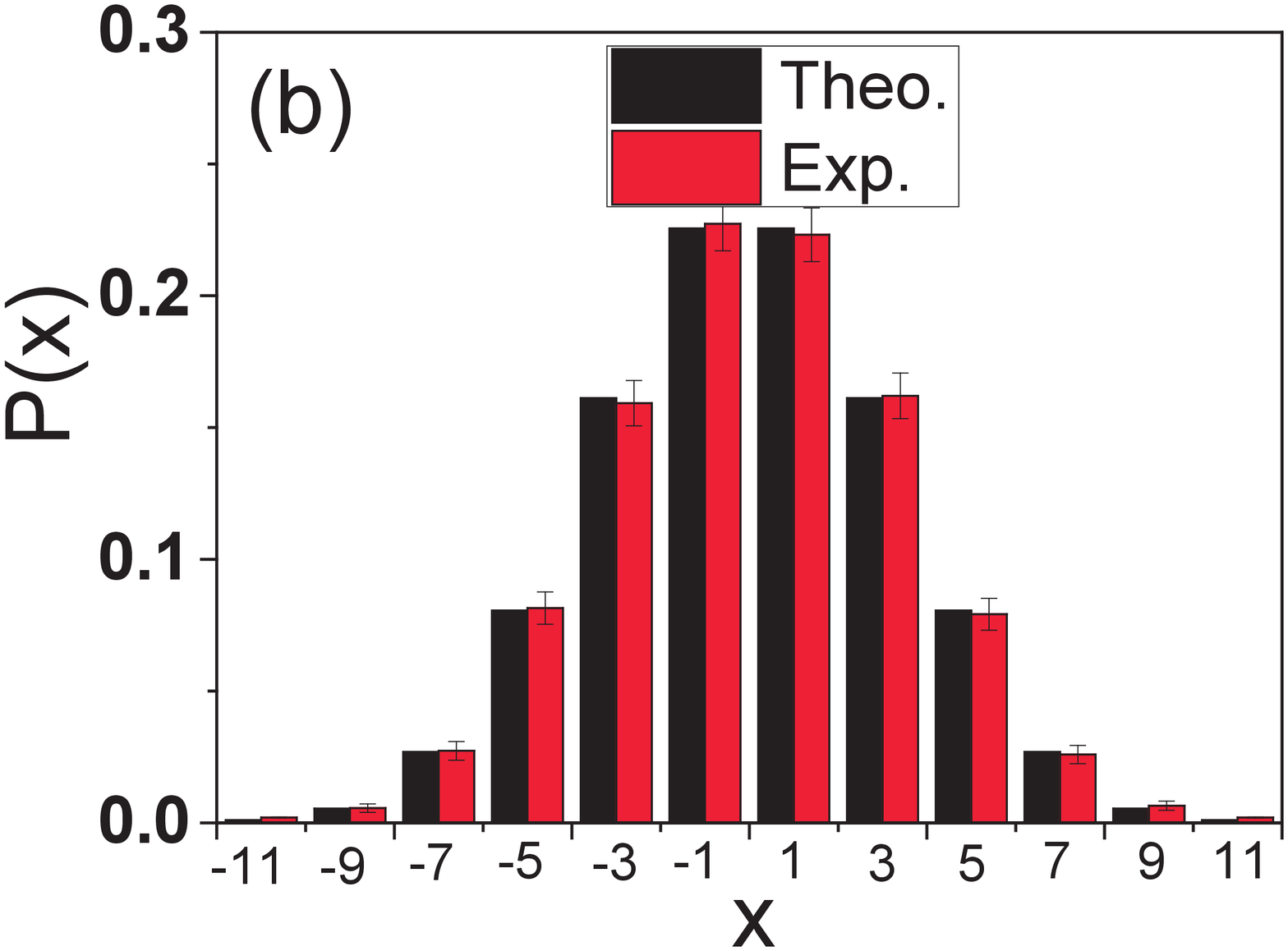}
    \includegraphics[width=0.48\textwidth]{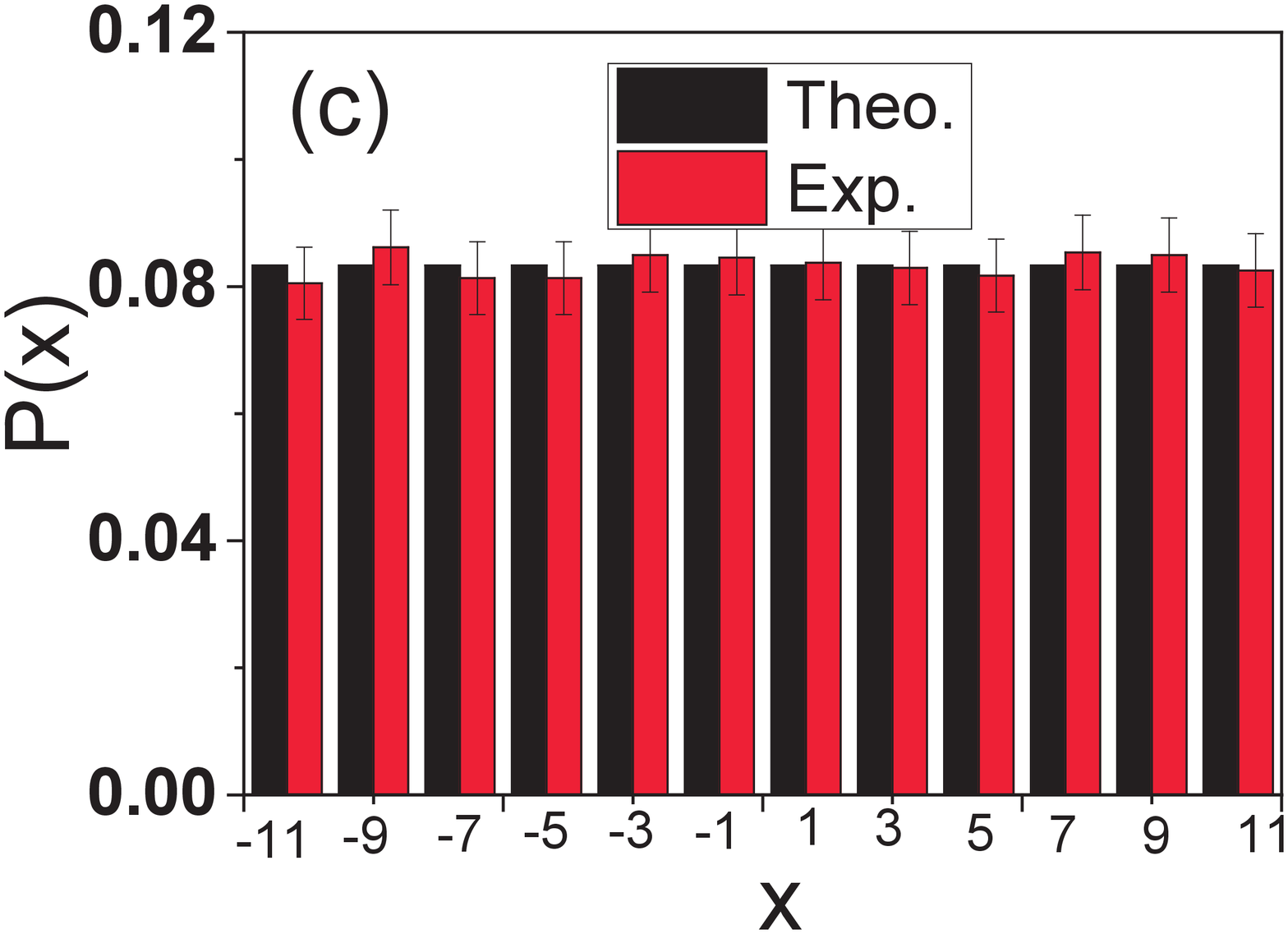}
    \caption{Experimental and theoretical position distribution $P(x)$ in step $11$.
    (a) Hadamard QW, (b) the Gaussian distribution, (c) the uniform distribution.
    Error bars are simulated from the statistical errors.
    }\label{Fig:distribution}
\end{figure}

The shift operation $\hat{S}_x$ is implemented in the time domain by an unbalanced Mach-Zehnder interferometer
composed with PBS2, PBS3 and reflected mirrors. The horizontal polarized photons traverse the optical network in $T = 72.9$~ns,
while the vertical ones take $\Delta t = 2.3$~ns longer.
The temporal difference $\Delta t$ correspond to a step in two positions $x \pm 1$.
After $t$-step evolution, the input light pulse is divided into $t+1$ time windows, i.e.,
the corresponding positions $x = -t, -t+2, \cdots, t$.

For detection, about one percent of photons are coupled out of the network and the most photons
go to the next round by a $99:1$ beam splitter (BS2).
The precise time, the number of photons, and the
polarization state of out-coupled photons are measured by QWP + HWP + PBS + SPD
(single-photon detector, Quantique id201).
The SPD is triggered by the electrical pulse from the AWG.
The nice timing and gated detection
allow us to monitor the whole evolution step by step with the reduced noise level.
The probability we detect the photon after a round-trip is $0.43$.
In order to obtain the position distribution of each step, we detect more than $10^4$ events in an overall measurement time.

\begin{figure}[t]
    \includegraphics[width=0.48\textwidth]{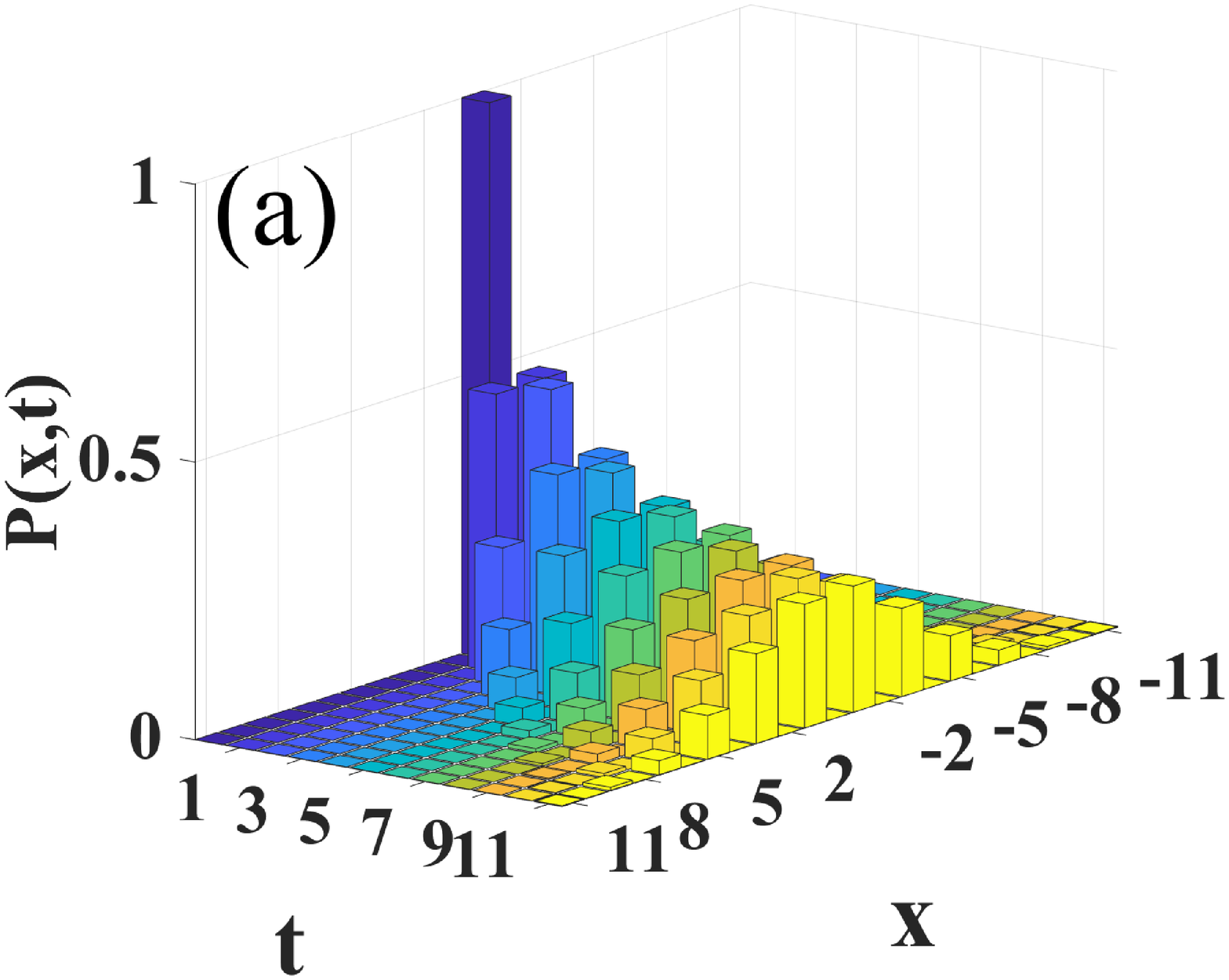}
    \includegraphics[width=0.48\textwidth]{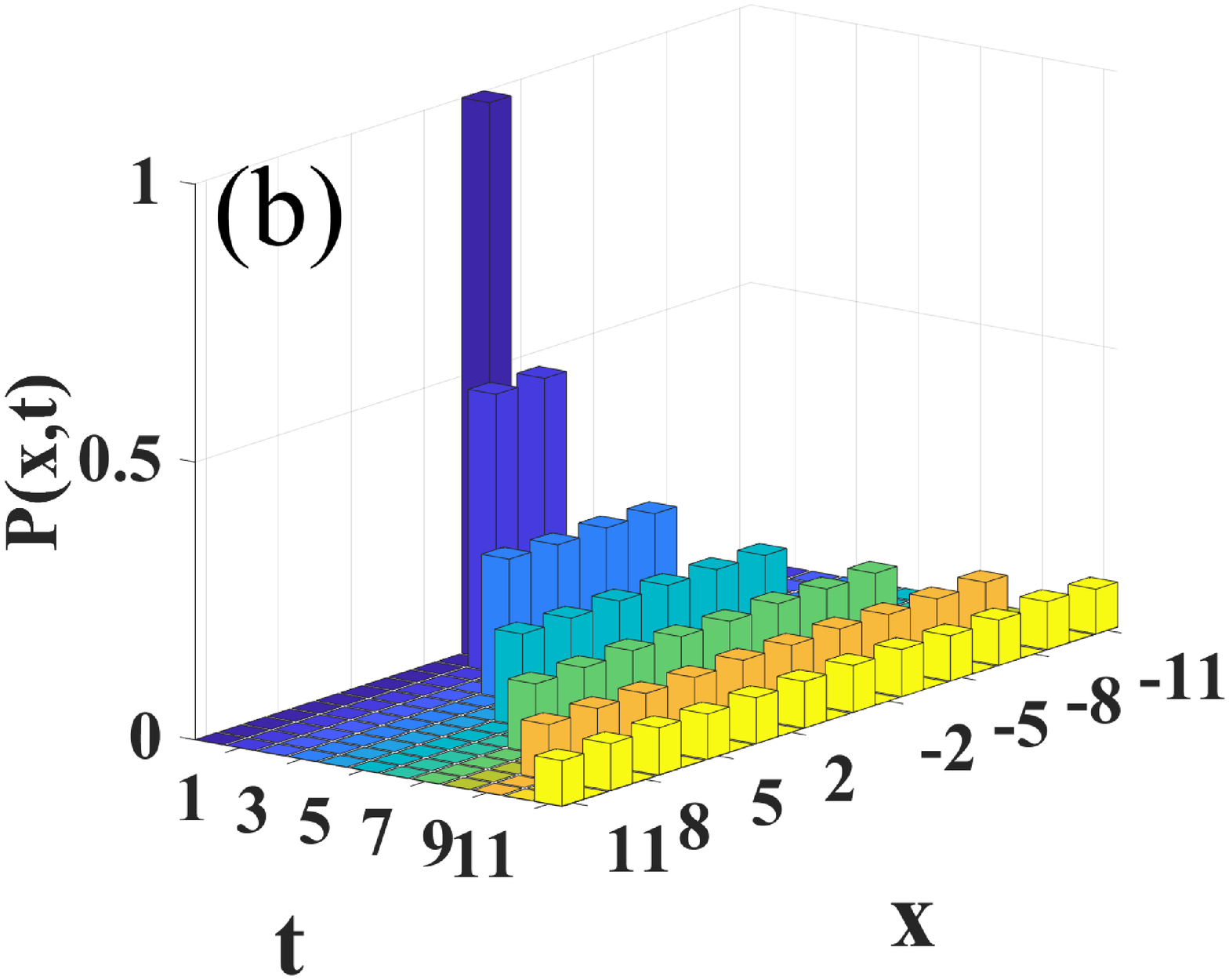}
    \caption{Experimental position distributions $P(x)$ in every odd steps from $1$ to $11$ step,
    (a) the Gaussian, (b) the uniform distribution. Results of even steps are similar with those of odd steps
     and not shown for clarity. Error bars are simulated from the statistical errors.
    }\label{Fig:distribution}
\end{figure}

\begin{table}[b]
\centering
\caption{The elements $|\rho^u_{x,x+2}|^2$ and $|\rho^u_{x,x}||\rho^u_{x+2,x+2}|$ of
the uniform distribution after $9$-step evolution. Error bars are estimated from the statistical errors.}
\label{Tab:01}
\setlength{\tabcolsep}{16pt}
\renewcommand\arraystretch{1.3}
\begin{tabular*}{8.6cm}{ccccc}\hline\hline
$x$ & $|\rho^u_{x,x+2}|^2$ & $|\rho^u_{x,x}||\rho^u_{x+2,x+2}|$   \\
\hline
$-9$  & $0.1754\pm0.0003$   &    $0.1754\pm0.0009$  \\
$-7$  & $0.0822\pm0.0003$   &    $0.0822\pm0.0011$  \\
$-5$  & $0.2443\pm0.0008$   &    $0.2443\pm0.0012$  \\
$-3$  & $0.2288\pm0.0008$   &    $0.2288\pm0.0012$ \\
$-1$   & $0.2200\pm0.0007$   &    $0.2200\pm0.0012$ \\
$1$    & $0.2419\pm0.0008$   &    $0.2419\pm0.0012$ \\
$3$    & $0.1735\pm0.0004$   &    $0.1735\pm0.0011$ \\
$5$    & $0.1659\pm0.0003$   &    $0.1659\pm0.0009$ \\
$7$    & $0.2499\pm0.0004$   &    $0.2499\pm0.0011$  \\
\hline
\end{tabular*}
\end{table}

To demonstrate the capability of generating versatile dynamics featured with arbitrary distributions in our setup,
we study three scenarios by programming different sets of coin operations.
We firstly implement the Hadamard QW to test the intrinsic coherence of our setup.
The initial state is prepared in a circular polarization $\psi(0)=\frac{1}{\sqrt{2}}(\ket{H}+i\ket{V})\ket{0}$.
And the coin operations at all positions for each step are the Hadamard operators
given in Eq.~(\ref{eq:C}) with $\theta_{x,t}=\pi/4$.
According to Eq.~(\ref{eq:Cexp}), the phase shifts
are $\phi_H(x,t)=\pi/4$ on H polarizations and $\phi_V(x,t)=3\pi/4$ on V polarizations. 
So the electrical pulses with the width $1$~ns and the height $0.127$~V are applied
on the EOM when every clockwise light pulses arrive and $0.392$~V for the arrival of all counter-clockwise light pulses.
The position distribution probabilities are obtained by adding up the the counts of
projection measurement on state $\ket{\text{H}}$ and $\ket{\text{V}}$ in each position,
and then normalizing them with the sum of all positions.
After $11$-step evolution, the position probabilities display the ballistic distribution
which is featured with two side peaks and the low probabilities around the original position as shown in Fig.~2a.
The performance of the evolution can be quantified by the similarity $F=\sum_x\sqrt{P_{Exp}P_{Theo}}$ between the measured position
distribution $P_{Exp}$ and the theoretical result $P_{Theo}$. It would be
$1$ provided that the experimental result was the same with the theoretical
prediction and $0$ for a totally mismatch. We obtain the feature of a completely coherent QW with the similarity $F = 0.988 \pm 0.002$.
Therefore, the nearly perfect coherent
evolution after $11$ steps in our setup is confirmed. 

\begin{table}[b]
\centering
\caption{The elements $|\rho^G_{x,x+2}|^2$ and $|\rho^G_{x,x}||\rho^G_{x+2,x+2}|$ of
the Gaussian distribution after $9$-step evolution. Error bars are calculated from the statistical errors.}
\label{Tab:02}
\setlength{\tabcolsep}{16pt}
\renewcommand\arraystretch{1.3}
\begin{tabular*}{8.6cm}{ccccc}\hline\hline
$x$ & $|\rho^G_{x,x+2}|^2$ & $|\rho^G_{x,x}||\rho^G_{x+2,x+2}|$   \\
\hline
$-5$  & $0.1414\pm0.0110$   &    $0.1414\pm0.0120$  \\
$-3$  & $0.1520\pm0.0013$   &    $0.1667\pm0.0016$ \\
$-1$   & $0.2000\pm0.0001$   &    $0.2293\pm0.0002$ \\
$1$    & $0.2027\pm0.00208$  &    $0.2272\pm0.0024$ \\
$3$    & $0.1342\pm0.0120$   &    $0.1342\pm0.0140$ \\
\hline
\end{tabular*}
\end{table}

We secondly realize the the coherent Gaussian distribution by using the QW with time- and position-dependent coin operations given by Eq.~(\ref{eq:C}), (\ref{eq:cosg}), (\ref{eq:sing}) and (\ref{eq:Cexp}).
The variable coin operations are realized by a precisely controlled phase shifts $\phi_{H/V}(x,t)$.
To generate the required phase shifts, the EOM is triggered by electrical pulses of the width $1$~ns and the time-bin-dependent voltages output from the AWG.
For example, the coin operation applied at position $x = 0$ for every even step is NOT gate, according to Eq.~(\ref{eq:Cexp}) the phase shifts on H polarizations $\phi_H(0,t)$ and V polarizations $\phi_V(0,t)$ are both $\pi/2$, so
we apply the electrical pulses with the height of $0.263$V on the EOM for the time bins $t_0=m(2T+\Delta t)$ $(m\in \mathbb{Z})$ and $t_0+\Delta t$ when the clockwise and counter-clockwise light pulses of $x=0$ arrive at the EOM.
The initial state can be arbitrary and here we use $\psi_G(0)=\ket{H}\ket{0}$.
We observe the Gaussian distributions after every steps.
The theoretical and experimental results with similarity $F = 0.995 \pm 0.003$ after $11$-
step are shown in Fig.~2b
and the measured distributions of every odd steps are shown in Fig.~3a.

Thirdly, the coherent uniform distribution is generated by employing the set of coin
operations expressed in Eq.~(\ref{eq:C}), (\ref{eq:cosu}) and (\ref{eq:sinu}). In a similar way,
the required phase shifts $\phi_{H/V}(x,t)$ are adjusted by applying
electrical pulses with the width $1$~ns and the time-bin-dependent voltages on the EOM.
We also use the initial state $\psi_u(0)=\ket{H}\ket{0}$. 
We measure the position distributions and obtain the uniform distribution after every steps from $1$ to $11$.
The theoretical and experimental results after $11$-step evolution with the similarity $F = 0.992 \pm 0.003$ are shown in Fig.~2c,
and the measured distributions of every odd steps are shown in Fig.~3b.
The deviations from the theoretical prediction are mainly caused by the small errors
on the rotation angle of the wave plates, the non-perfect separation of H
and V polarizations of the PBS, and the slight different losses between the long and
short paths in the unbalanced MZ interferometer.

To prove that the multi-path coherence is well preserved during the whole evolution,
it is necessary to certify that the evolved states of uniform and Gaussian distributions are pure.
By employing the method of Ref.~\cite{GD15}, the multi-path state is pure if
$\left|\rho_{x,x+2}\right|^2=\left|\rho_{x,x}\right|\left|\rho_{x+2,x+2}\right|$
holds for all $x=-t, -t+2, \cdots, t-2$.
So we perpetuate a step of the quantum walk following the above $t$ steps evolution
and then transform the tomography on the two neighboring positions bases
into the polarization bases.
For each pair light pules corresponding with states $\{\ket{H}\ket{x},\ket{H}\ket{x+2}\}$ with $x=-t, -t+2, \cdots, t-2$, firstly,
the NOT operation is operated on the polarization part of $\ket{H}\ket{x}$
and identity operator on $\ket{H}\ket{x+2}$,
so $\ket{V}\ket{x}$ and $\ket{H}\ket{x+2}$ are obtained.
Then the shift operation is utilized to combine
the light pulse of $\ket{V}\ket{x}$ and that of $\ket{H}\ket{x+2}$ into one light pulse. Finally,
by state tomography on the polarization bases $\{\ket{H},\ket{V},\ket{H}+\ket{V},\ket{H}-i\ket{V}\}$,
$\left|\rho_{x,x+2}\right|^2=\left|\rho_{x,x}\right|\left|\rho_{x+2,x+2}\right|$ can be proved if the corresponding $\left|\rho_{H,V}\right|^2=\left|\rho_{H,H}\right|\left|\rho_{V,V}\right|$ works.
Since the local unitary operation cannot change the purity of the state,
the influence on the polarization from the single-mode fiber is neglected.

We take the evolved state after $9$-step evolution as an example
because the intensity after $11$-step dynamics
is too low to execute the precise state tomography due to the limited laser light intensity and the high loss of our optical network.
Based on the experimental measurement results of the uniform distribution,
we obtain the values of $\left|\rho^u_{x,x+2}\right|^2$ and
$\left|\rho^u_{x,x}\right|\left|\rho^u_{x+2,x+2}\right|$
for each $x=-9, -7, \cdots, 7$ which are listed in Table.~1,
And the partial entries $\left|\rho^G_{x,x+2}\right|^2$ and
$\left|\rho^G_{x,x}\right|\left|\rho^G_{x+2,x+2}\right|$ for $x=-5, -3, \cdots, 5$
of Gaussian distribution are obtained as shown in Table.~2,
because the intensities in Gaussian case of neighboring sites $\ket{-9}$ and $\ket{-7}$
are too weak to apply precise measurement on the polarization compared with the uniform distribution.
The experimental results show that the equation
$\left|\rho_{x,x+2}\right|^2=\left|\rho_{x,x}\right|\left|\rho_{x+2,x+2}\right|$
works and so we confirm that the evolved walker states are pure.
Therefore, it is concluded here that the coherence is well preserved during the dynamic evolutions.

\begin{figure}[t]
    \includegraphics[width=0.48\textwidth]{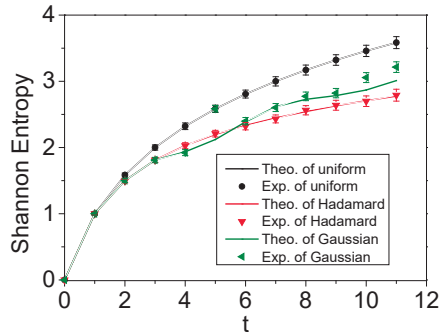}
    \caption{ Intrinsic randomness in coherent position states of Hadamard, Gaussian and uniform distribution
    as a function of the evolution step $t$.
        The error bars are simulated from the statistical errors.
    }\label{Fig:distribution}
\end{figure}

We has experimentally proved that the coherent superposition 
of many position states is well preserved, so it can be employed as an intrinsic random
resource with the unpredictability~\cite{SC19}. The randomness can be quantified
by the Shannon entropy $R(\ket{\psi}\bra{\psi})=-\sum_x P_x log_2 P_x$, in which the position probability
$\{P_x\}$ had been obtained experimentally. The Shannon entropy of the uniform, Gaussian and Hadamard
distributions as a function of evolution time are shown in Fig.~4. It is evident that the
randomness can be significantly enhanced by increasing the evolution steps $t$. While for the same $t$,
the coherent uniform distribution reach the maximal randomness compared with Gaussian and
ballistic distributions. The random sequence can be generated directly
through the coherent uniform distribution after a single quantum measurement
and the bit rates can be enhanced by increasing the evolution steps.

\section{Conclusion}
We have realized the time- and position-inhomogeneous discrete-time quantum walk (QW)
by employing a flexible optical loop architecture.
The rich dynamics can be obtained
by the stably and precisely controlling on the time- and position-dependent coin operations.
We study the quantum walker suffered three different unitary evolutions.
Firstly, we implement the Hadamard QW to obtain the ballistic distribution
which confirms a nearly perfect coherence of our setup.
Secondly, we program the time- and position-dependent coin operations
by applying electrical pulses with the proper pattern on the EOM to obtain the coherent Gauss distribution,
in which the position distribution is the same with CRW, meanwhile the fairly good quantum interference is maintained.
Finally, electrical pulses with another different pattern are applied to generate the coherent uniform superposition of position states.
And the coherence is certified by the purity of the evolved walker state.
Besides, the coherent uniform state is the high-order W-type entanglement and can
be used to produce the genuine multi-bit random numbers. The bit rate can be readily
improved by increasing the evolution steps and fully
meet the requirement of security by exploiting the intrinsic uncertainty of the quantum state.
We provide an accurately controlled platform to generate arbitrarily desired dynamics featured
with variable position distributions, and consequently pave the way for the realization of quantum-walk-based quantum computation,
simulations and quantum information protocols.
And our method has the advantage on the scalability and can be extended to multi-partite
and the high-dimension QW, which may exhibit more rich dynamics.

\acknowledgements
This work was supported by National Key R\&D Program of China (No.
2019YFA0705000 and 2019YFA0308700), the Key R\&D Program of Guangdong Province (Grant No.
2018B030329001), National Natural Science Foundation of China (Grants No. 51890861, No. 11547031,
No. 11705096  No. 11674169, No. 11627810, No. 61705033, No. 11690031, and  No. 11974178).


\end{document}